\documentclass[a4paper]{article}

\usepackage{graphicx}
\usepackage{longtable}

\usepackage{amssymb}
\usepackage{amsmath}
\usepackage{hyperref}
\usepackage{authblk}
\title{ Magnetic Field Focusing of Hyperfine Interaction in Hydrogen}

\author[1,2]{M.A. Andreichikov \thanks{andreichicov@mail.ru}}

\author[1,2]{B.O. Kerbikov \thanks{borisk@itep.ru}}

\author[1]{Yu.A. Simonov \thanks{simonov@itep.ru}}

\affil[1]{Institute of Theoretical and Experimantal Physics\\
117118, Moscow, B. Cheremushkinskaya 25, Russia}
\affil[2]{Moscow Institute of Physics and Technology\\
141701, Moscow region, Dolgoprudny, Institutsky lane 9, Russia}

\begin{document}

\newcommand{\vect}[1]{\mathbf{#1}}
\newcommand{\gvect}[1]{\boldsymbol{#1}}
\newcommand{\dif}[2]{\frac{\mathrm{d}{#1}}{\mathrm{d}{#2}}}
\newcommand{\ddif}[2]{\frac{\mathrm{d}^2{#1}}{\mathrm{d}{#2}^2}}
\newcommand{\pdif}[2]{\frac{\partial{#1}}{\partial{#2}}}
\newcommand{\pddif}[2]{\frac{\partial^2{#1}}{\partial{#2}^2}}

\date{\today}

\maketitle

\begin{abstract}

We find a new correction to hyperfine splitting in the ground state of hydrogen atom in magnetic field. The physical basis for this effect is the reduction of the size of the electron orbit in magnetic field. As a result, the value of the wave function at the origin increases which can be called magnetic focusing. Another magnetic field induced effect is the appearance of field dependent tensor forces.

\end{abstract}

\section{}

The spectrum of hydrogen atom (HA) in strong magnetic field (MF) was found long ago \cite{1} and is presented in textbooks \cite{2}. In recent years we are witnessing the rise of interest to this subject. This is probably due to the fact that huge MF up to $eB \sim \Lambda_{QCD}^2 \sim 10^{19} \ G$ has become a physical reality. Such field is created (for a short time) in heavy ion collisions at RHIC and LHC \cite{3}. The field about four orders of magnitude less is anticipated to operate in magnetars \cite{4}. Several interesting MF induced effects in QCD are under investigation now both from theoretical and experimantal sides \cite{5}. Among new results in physics of HA in MF necessary to mention the conclusion that in superstrong MF radiative corrections screen the Coulomb potential thus preventing the ``fall to the center'' phenomenon to occur \cite{6}. In the present paper we discuss another MF induced effect, namely MF focusing of hyperfine interaction. The HA is squeezed by MF and the value of the wave function at the origin increases. This changes the Hamiltonian of hyperfine interaction. In addition, in MF the HA takes the form of an elongated ellipsoid. As a result field dependent tensor forces are induced. Experimentally magnetic focusing manifests itself as an additional shift of hyperfine states on top of the standard Zeeman splitting.

\section{}

We begin by introducing the units to be used and reminding some basic equations. We put $\hbar = c = 1$, $\alpha = e^2 = 1/137$, dimensionless MF is defined as $H = B/B_a$, $B_a = m^2e^3 = 2.35 \cdot 10^9 \ G$ is the so-called atomic MF. At $B = B_a$ the Bohr radius $a_B = (\alpha m)^{-1}$ becomes equal to the magnetic, or Landau, radius $a_H = (eB)^{-1/2}$, the oscillator energy $\omega = eB/2m$ becomes equal to Rydberg energy $Ry = m\alpha^2/2$. In this system of units $GeV^2 = 1.45 \cdot 10^{19} \ G$. 

The problem of HA in uniform MF is convenient to solve in cylindrical coordinates $(\gvect{\rho},z)$ using the London gauge $\vect{A} = \frac{1}{2} \vect{B} \times \vect{r}$, hence $\vect{B}$ is directed along the z-axis. The nonrelativistic Hamiltonian reads

\begin{equation}
  \label{eq:1}
  \hat{\mathcal{H}} = - \frac{1}{2m} \left( \Delta_{\perp} + \pddif{}{z} \right) + \omega \hat{l}_z + \frac{m \omega^2 \rho^2}{2} - \frac{\alpha}{\sqrt{\rho^2 + z^2}} + \mu_B \sigma_z B,
\end{equation}

where $\Delta_{\perp}$ is the transverse part of the Laplacian, $\mu_B = e/2m$, $e$ is the absolute value of the electron charge, $\gvect{\sigma} = 2\vect{s}_e$, $\sigma_z = \pm 1$. The Schrodinger equation described by the Hamiltonian (\ref{eq:1}) does not allow the separation of the coordinates $\gvect{\rho}$ and $z$. However in superstrong MF limit $H \gg 1$ the ``fast'' MF variable $\gvect{\rho}$ and the ``slow'' Coulomb variable $z$ may be separated in the form of the adiabatic ansatz \cite{7,1}.
\begin{equation}
  \label{eq:2}
  \Psi(\gvect{\rho},z) = R_{n_{\rho}m}(\gvect{\rho})\chi_{n n_{\rho}m}(z)\chi_{\sigma_z}.
\end{equation}
For $H \gg 1$ the dominant role is played by the lowest Landau level (LLL) with $n_{\rho} = 0$, $m = 0,-1,-2,...$, $\sigma_z = -1$. For this state the energy of the oscillations in $\gvect{\rho}$-plane and the spin magnetic energy $\mu_BB$ compensate each other. Finally we note that electron becomes relativistic for MF larger than the Schwinger one $B_c = m^2/e$ with only one exception: electron at LLL remains non-relativistic \cite{6}.

\section{}

Now we come to the subject of the paper. Hyperfine splitting (hfs) in the ground state of HA is measured to 13 significant figures in frequency units \cite{8,9} 
\begin{equation}
  \label{eq:3}
  \Delta E_{hfs} = 1420.4057517667(9) \ MHz,
\end{equation}
which amounts approximately to $5.9 \cdot 10^{-6} \ eV$. It corresponds to the 21cm  line discovered in 1951 \cite{10} and since then thought to be primary tool in radioastronomy. The hfs can be found to lowest order in $\alpha$ from  Breit magnetic dipole interaction
\begin{equation}
  \label{eq:4}
  \hat{\mathcal{H}}^{(0)}_{hf} = \frac{8\pi}{3} g_p \mu_B \mu_N  (\gvect{\sigma}_e \cdot \gvect{\sigma}_p) \delta(\vect{r}),
\end{equation}
where $g_p = 2.79$, $\mu_N = e/2m_p$, and the superscript signifies the absence of external MF. The first order perturbation of $\mathcal{\hat{H}}_{hf}^{(0)}$ gives
\begin{equation}
  \label{eq:5}
  \Delta E_{hfs} = \frac{32\pi}{3}g_p \mu_B \mu_N |\Psi(0)|^2.
\end{equation}
There are three types of corrections to this expression: a)relativistic effects, b) QED, and c) nuclear structure. They have been thoroughly discussed in the literature - see e.g., \cite{8,9}.

With MF imposed equations (\ref{eq:4}) and (\ref{eq:5}) experience important changes. The problem is considered in detail in textbooks on quantum mechanics \cite{2}. Our solution contains two new points, namely magnetic focusing and the presence of tensor forces. Both effects are caused by the action of MF which enhances the wave function at the origin and gives a non-spherical form to the HA. To get the needed expression for $\hat{\mathcal{H}}_{hf}$ we start from the Biot-Savart law \cite{2,11}. The operator $\hat{\mathcal{H}}_{hf}$ has the form $\hat{\mathcal{H}}_{hf} = -g \mu_N (\gvect{\sigma}_p \vect{B}')$, where $\vect{B}'$ is the MF created at the origin by the spin part of the electron current. For the current one has 
\begin{equation}
  \label{eq:6}
  \vect{j}_e = -\mu_B\nabla \Psi^2 \times \gvect{\sigma}_e, 
\end{equation}
where the function $\Psi(\gvect{\rho},z)$ is real and $\varphi$-independent since we consider the ground state with $l_z=0$. External MF $\vect{B}$ enters via the wave function $\Psi$ to be specified below. Next we have 
\begin{equation}
  \label{eq:7}
  \vect{B}' = \int dV  \frac{\vect{n} \times \vect{j}}{r^2},
\end{equation}
\begin{equation}
  \label{eq:8}
  \vect{n} \times \vect{j} = -\mu_B \left[ \nabla \Psi^2(\vect{n} \cdot \gvect{\sigma}_e) - \gvect{\sigma}_e(\vect{n} \cdot \nabla \Psi^2) \right],
\end{equation}
with $\vect{n}$ being the unit vector along the line connecting $dV$ and the origin where proton is placed. From ({\ref{eq:7})-(\ref{eq:8}) one obtains

\begin{equation}
  \label{eq:12}
\hat{\mathcal{H}}_{hf}   = g \mu_B \mu_N \left[ \int dV \frac{(\gvect{\sigma}_p \cdot \nabla \Psi^2)(\gvect{\sigma}_e \cdot \vect{r})}{r^3} - (\gvect{\sigma}_e \cdot \gvect{\sigma}_p) \int dV \frac{(\vect{r} \cdot \nabla \Psi^2)}{r^3} \right].
\end{equation}
Integrating by parts one can easily convert (\ref{eq:12}) into the standard form
\begin{equation}
      \label{eq:13}
 \hat{\mathcal{H}}_{hf}  = g \mu_B \mu_N  \int dV \Psi^2 \left[ \frac{8\pi}{3} (\gvect{\sigma}_p \cdot \gvect{\sigma}_e)\delta(\vect{r}) + 
 \frac{3(\gvect{\sigma}_p \cdot \vect{r})(\gvect{\sigma}_p \cdot \vect{r}) - (\gvect{\sigma}_p \cdot \gvect{\sigma}_e)r^2}{r^5} \right] 
\end{equation}
The form (\ref{eq:12}) does not explicitly contain the $\delta$-function and is therefore better suited for calculations. To proceed further, we need an explicit expression for the wave function $\Psi$.

\section{}

Attempts to find eigenvalues and eigenfunctions of Hamiltonian (\ref{eq:1}) have a long history - see \cite{12} for a list of references. We use a variational method as many authors listed in \cite{12} did. In certain features our approach bears a resemblance to that of \cite{13,14}. The wave function for the ground state is written as 
\begin{equation}
  \label{eq:9}
  \Psi_0(\rho,z) = \sqrt{N} \exp \left(-\frac{\rho^2}{2r_{\perp}^2} - \frac{z^2}{2r_z^2} \right) ,
\end{equation}
where $N =(\pi^{3/2}r_{\perp}^2r_z)^{-1}$. The two parameters $r_{\perp}$ and $r_z$ are fitted at each value of $H$. According to \cite{15} $r_{\perp} \sim (H)^{-1/2}$, $r_z \sim (\ln H)^{-1}$ The rationale for choosing $\Psi_0$ in the form (\ref{eq:9}) is the following: a) it has a desired form of an elongated ellipsoid, b) it has an axial symmetry and invariant under reflection with respect to $\gvect{\rho}$-plane, c) our calculations show that for $H \gg 1$ the fitted wave function (\ref{eq:9}) is close to that obtained in \cite{6,15}, and for $0 < H < 1$ the results are in agreement with very accurate calculations of several authors, e.g., \cite{16}. Necessary to stress that by taking the trial wave function in a simple form (\ref{eq:9}), we shall be able to expose very clearly the new contribution into the hyperfine splitting. The new effect is independent on the concrete form of the trial wave function, or, more generally, on the method to solve the problem.

The ground state energy is defined from 
\begin{equation}
  \label{eq:10}
  E_o = \langle \Psi_0 | \hat{\mathcal{H}}_0 | \Psi_0 \rangle, \ \pdif{E_0}{r_{\perp}} = 0,\ \pdif{E_0}{r_z} = 0,
\end{equation}
where $\hat{\mathcal{H}}_0$ is obtained from (\ref{eq:1}) by removing the term $\mu_B \sigma_z B$.
Straightforward calculation yields the following result for $E_0$ 
\begin{equation}
  \label{eq:11}
 E_0(r_{\perp},r_z) = \frac{1}{2mr_{\perp}^2}\left(1 + \frac{\beta^2}{2}\right) + \frac{m \omega^2 r_{\perp}^2}{2} - \frac{\alpha \beta}{r_{\perp} \sqrt{\pi(1-\beta^2)}} \ln \frac{1+\sqrt{1-\beta^2}}{1-\sqrt{1-\beta^2}},
\end{equation}
where $\beta = r_{\perp}/r_z < 1$ for $B > 0$. Minimization of (\ref{eq:11}) according to (\ref{eq:10}) yields $r_{\perp}$ and $r_z$ as functions of $H$. For illustrative purposes consider two limiting cases: a)$H = 0$, then $\omega = 0$, $r_{\perp} = r_z$, $E_0 = 4m\alpha^2/3\pi \simeq 0.85 \ Ry$ in line with \cite{17}, b) free particle in MF, then we obtain $r^2 = (m\omega)^{-1}$, $E_0 = \omega$.

In Fig.\ref{fig:1} we plot the energy $E_0$ as a function of $H$ in comparison with the results of \cite{16}. The deviation from the elaborated calculation \cite{16} does not exceed $15 \%$. In Fig.\ref{fig:2} we display the radii $r_{\perp}$ and $r_z$ as functions of $H$. This figure demonstrates how the deformation of the wave function with $H$ proceeds. 

\begin{figure}[h!]
  \centering
  \includegraphics[width=0.8\textwidth]{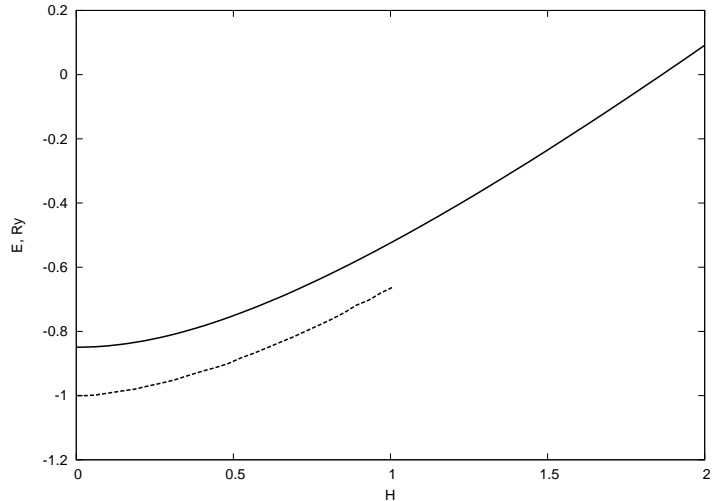}
  \caption{Plot of $E_0$ (without spin contribution) vs. $H$. Solid curve - present calculation, dashed one from \cite{16}}
  \label{fig:1}
\end{figure}

\begin{figure}[h!]
  \centering
  \includegraphics[width=0.8\textwidth]{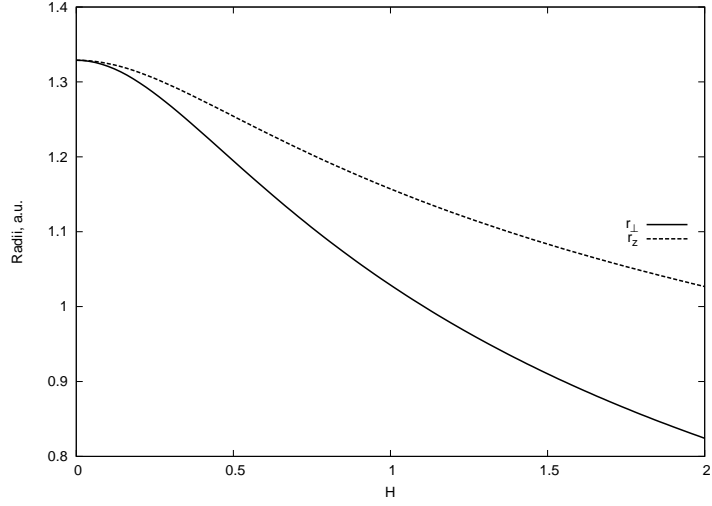}
  \caption{The radii $r_{\perp}$ (solid line) and $r_z$ (dashed line)in atomic units as functions of $H$.}
  \label{fig:2}
\end{figure}

\section{}

With the fitted wave function at our disposal we return to (\ref{eq:12}) perform the integration. 

The integrals can be evaluated analytically with the following final result:
\begin{equation}
  \label{eq:13}
  \hat{\mathcal{H}}_{hf} = g \mu_B \mu_N \left[ (F_1(H) + F_2(H))(\gvect{\sigma}_e \cdot \gvect{\sigma}_p) + (F_1(H) - F_2(H))\sigma_{pz}\sigma_{ez} \right],
\end{equation}
where 
\begin{equation}
  \label{eq:14}
  F_1(H) = \frac{1}{\sqrt{\pi}r_{\perp}^2r_z}\left[\frac{2}{1-\beta^2} - \frac{\beta^2}{(1-\beta^2)^{3/2}} \ln \frac{1+\sqrt{1-\beta^2}}{1-\sqrt{1-\beta^2}} \right],
\end{equation}
\begin{equation}
  \label{eq:15}
  F_2(H) = \frac{2}{\sqrt{\pi}r_z^3}\left[-\frac{2}{1-\beta^2} + \frac{1}{(1-\beta^2)^{3/2}} \ln \frac{1+\sqrt{1-\beta^2}}{1-\sqrt{1-\beta^2}} \right].
\end{equation}
 At $H \rightarrow 0$, $\beta \rightarrow 1$, $r_{\perp} = r_z = r$, and from (\ref{eq:14}) - (\ref{eq:15}) one obtains 
\begin{equation}
  \label{eq:16}
  F_1 = F_2 = F = \frac{4}{3\sqrt{\pi}}r^{-3} = \frac{4\pi}{3}|\Psi(0)|^2,
\end{equation}
and we return to Eqs. (\ref{eq:4}) - (\ref{eq:5}). At $H \gg 1$ we have
\begin{equation}
  \label{eq:17}
  \beta \sim \frac{\ln H}{\sqrt{H}},\ F_1 \sim H \ln H, \ F_2 \sim \sqrt{H}\ln^2 H.
\end{equation}
Equations (\ref{eq:13})-(\ref{eq:16}) comprise the essence of the physical process which can be called ``Magnetic Focusing of Hyperfine Interaction''. MF compresses the HA thus increasing the wave function of the origin and giving rise to MF-dependent tensor component. 

\section{}

The next task is to see how our results modify the standard Zeeman splitting effect. 

In MF the ground state of HA is splitted into four levels with their energies obtained by the diagonalization of the Hamiltonian

\begin{equation}
  \label{eq:18}
  \hat{\mathcal{H}}_{hf}' = \hat{\mathcal{H}}_{hf} + \mu_B (\gvect{\sigma}_e  \cdot \vect{B}) - g \mu_N (\gvect{\sigma}_p \cdot \vect{B}),
\end{equation}
where $\hat{\mathcal{H}}_{hf}$ is given by (\ref{eq:13}) and has two new important features. First, it depends on MF since the parameters $r_{\perp}$ and $r_z$ entering into $F_1$ and $F_2$ are fitted at each value of $H$. Physically, this is tantamount to focusing of HA wave function at the origin. Second, $\hat{\mathcal{H}}_{hf}$ contains the term proportional to $\sigma_{pz}\sigma_{ez}$ reflecting the deviation of HA from spherical symmetry. Let us focus on the transitions between the states which at $B = 0$ correspond to $|a \rangle = |S=1, S_z = 0 \rangle$ and $|b \rangle = |S=0, S_z = 0 \rangle$. From (\ref{eq:18}) one gets 
\begin{equation}
  \label{eq:19}
\nu = E_a - E_b = \Delta E_{hfs} \sqrt{ \gamma^2 + \left(\frac{2\mu_BB}{\Delta E_{hfs}}\right)^2 \left(1 + g\frac{m}{m_p}\right)^2 },
\end{equation}
where $\Delta E_{hfs}$ is given by (\ref{eq:5}) and $\gamma = (F_1 + F_2)/2F$. Without Magnetic Focusing  $\gamma =1$ and the standard expression is retrieved. The quantity of interest is the difference $\delta \nu = \nu - \nu_0$ with $\nu_0$ corresponding to $\gamma=1$. Here we present estimates of $\delta \nu$ in the two limiting regimes of super-strong ($H \gg 1$) and weak ($H \ll 10^{-7}$) MF. Performing simple calculations starting from (\ref{eq:14}), (\ref{eq:15}) and (\ref{eq:19}) we arrive at the following results:
\begin{equation}
  \label{eq:20}
  \delta \nu \simeq \alpha^6 \left(\frac{m}{m_p}\right) m (H \ln^2 H) \simeq 10^{-6}(H \ln^2 H) \ MHz
\end{equation}
for $H \gg 1$ and 
\begin{equation}
  \label{eq:21}
  \delta \nu \simeq \Delta E_{hfs} \left(1 - \frac{r_{\perp}^2}{r_z^2} \right)
\end{equation}
for $H \ll \alpha^2 \frac{m}{m_p} \simeq 10^{-7} \simeq 100 \ G$. Evaluation of the quantity $r_{\perp}^2/r_z^2$ in the weak field limit requires very accurate numerical calculations which will be presented in the forthcoming publication. We remind that the present hydrogen maser experiments are sensetive to the variations of the Zeeman splitting of the order of $1 \ mHz$ \cite{18}. In Fig.\ref{fig:3} we show $\delta \nu$ in a rather wide interval of $H$. The growth of $\delta \nu$ with $H$ reflects the gradual deviation of HA from the spherical symmetry.

\begin{figure}[h!]
  \centering
  \includegraphics[width=0.8\textwidth]{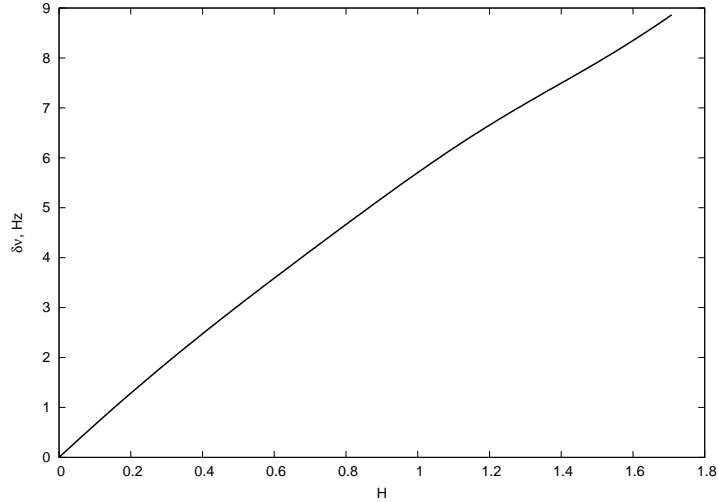}
  \caption{The frequency shift $\delta \nu$ (see the text) vs. $H$.}
  \label{fig:3}
\end{figure}
\section{}
Magnetic Field Focusing considered here for the HA is a universal phenomenon important for any quantum system/reaction in presence of MF as soon as the wave function at the origin is an important parameter. In particular, it leads to the modification of $\beta$-decay rate in MF \cite{19}. Another example is the spectrum of quark-antiquark system \cite{20}. Interesting effects occur also in super-strong MF created in heavy-ion collisions. A few words are needed to add concerning related problems left beyond the scope of the present paper. Magnetic focusing in muonic hydrogen may be easier to observe experimentally \cite{9}. For $H \gg 1$ another correction comes into play - proton can not be considered as infinitely heavy and problem becomes a two-body one \cite{21}. 

The authors are grateful for many useful discussions and remarks to M.I. Vysotsky, S.I. Godunov, V.S. Popov and M.E. Eides.


\begin{thebibliography}{99}

\bibitem{1} R.J. Elliott and R. Loudon, {\rm J. Phys. Chem. Sd.}, {\bf 15}, 196 (1960)
\bibitem{2} L.D. Landau and E.M. Lifshitz, {\rm Quantum Mechanics. Course of Theoretical Physics, vol.3}, Pergamon Press, Oxford (1978)
\bibitem{3} D.E. Kharzeev, L.D. McLerran and H.J. Warringa, {\rm Nucl. Phys}, {\bf A803}, 227 (2008); V.S. Skokov, A. Illarionov and V. Toneev, {\rm Int. J. Mod. Phys.} {\bf A24}, 5925 (2009)
\bibitem{4} C. Kouveliotou, R.C. Duncan and C. Thompson, {\rm Sci. Am.}, {\bf 228N2}, 24 (2003); A. Harding and Dong Lai, {\rm Rept. Prog. Phys.}, {\bf 69}, 2631 (2006).
\bibitem{5} B.O. Kerbikov and M.A. Andreichikov, arXiv:1211.1937, {\rm Contribution to the Proceedings of ``QUARKS 2012'' International Seminar}, Yaroslavl, June 4-10, 2012.
\bibitem{6} A.E. Shabad and V.V. Usov, {\rm Phys. Rev. Lett.}, {\bf 98}, 180403 (2007); {\rm Phys. Rev.} {\bf D73}:125021 (2006); 
B. Machet and M.I. Vysotsky, {\rm Phys. Rev.}, {\bf D83}:025022 (2011); S.I. Godunov, B. Machet and M.I. Vysotsky, {\rm Phys. Rev.} {\bf D85}:044058 (2012); S.I. Godunov and M.I. Vysotsky, arXiv:1304.7940.
\bibitem{7} L.I. Schiff and H. Snyder, {\rm Phys. Rev.} {\bf 55}, 59 (1939)
\bibitem{8} S.G. Karshenboim, {\rm Phys. Rept.} {\bf 422}, 1 (2005)
\bibitem{9} M.I. Eides, H. Grotch and V.A. Shelyuto, {\rm Phys. Rept.} {\bf 342}, 63 (2001)
\bibitem{10} H.I. Ewen and E.M. Purcell, {\rm Nature} {\bf 168}, 356 (1951)



\bibitem{11} E. Fermi, {\rm Z. Phys.} {\bf 60}, 320 (1930)
\bibitem{12} H. Friedrich and D. Wintgen, {\rm Phys. Rept.} {\bf 183}, 37 (1989)
\bibitem{13} R. Cohen, J. Lodenquai and M. Ruderman, {\rm Phys. Rev. Lett}, {\bf 25}, 467 (1970).
\bibitem{14} M. Bachmann, H. Kleinert and A. Pelster, {\rm Phys. Rev.}, {\bf A62}, 052509/1-21 (2000).
\bibitem{15} B.M. Karnakov and V.S. Popov, {\rm J. Exp. Theor. Phys.} {\bf 97}, 890 (2003); {\rm Zh. Eksp. Teor. Fiz.}, {\bf 141}, 5 (2012).
\bibitem{16} H.C. Praddaude, {\rm Phys. Rev.} {\bf A6}, 1321 (1972)
\bibitem{17} R.P. Feynman and H. Kleinert, {\rm Phys. Rev.} {\bf A34}, 5080 (1986).
\bibitem{18} M.A. Humphrey et. al., {\rm Phys. Rev.} {\bf A68}, 063807 (2003); D.F. Phillips et. al., {\rm Phys. Rev.} {\bf D63}, 111101 (2001)
\bibitem{19} K.A. Kouzakov and A.I. Studenikin, {\rm Phys. Rev.} {\bf C72}, 015502 (2005)
\bibitem{20} M.A. Andreichikov, B.O. Kerbikov, V.D. Orlovsky and Yu.A. Simonov, {\rm Phys. Rev.} {\bf D87}, 094029 (2013); Yu.A. Simonov, arXiv:1304:0365; C.S. Machado et. al., arXiv:1305.3308 
\bibitem{21} H. Herold, H. Ruder and G.Wunner, {\rm J.Phys.} {\bf B14}, 751 (1981)

\end{thebibliography}
\end{document}